\documentclass{kapedbk}
\begin{document}
\baselineskip22pt

\normallatexbib

\articletitle{The Hypothesis of Locality and its Limitations}

\author{Bahram Mashhoon}
\affil{Department of Physics and Astronomy \\
University of Missouri-Columbia\\ Columbia, Missouri 65211, USA}

\begin{abstract}
The hypothesis of locality, its origin and consequences are discussed. 
This supposition is necessary for establishing the local spacetime frame
of accelerated observers; in this connection, the measurement of length
in a rotating system is considered in detail.  Various limitations of the
hypothesis of locality are examined.  
\end{abstract}

\section{Introduction}
The basic laws of microphysics have been formulated with respect to ideal
inertial observers.  However, all actual observers are accelerated.  To
interpret the results of experiments, it is therefore necessary to
establish a connection between actual and inertial observers.  This is
achieved in the standard theory of relativity by means of the hypothesis
of locality, namely, the assumption that an accelerated observer at each
instant along its worldline is physically equivalent to an otherwise
identical momentarily comoving inertial observer.  In this way a
noninertial observer passes through a continuous infinity of hypothetical
momentarily comoving inertial observers \cite{[1]}.  

The hypothesis of locality stems from Newtonian mechanics, where the
state of a particle is given at each instant of time by its position and
velocity.  Thus the accelerated observer and the hypothetical inertial
observer share the same state and are therefore equivalent.  Hence, the
treatment of accelerated systems in Newtonian mechanics requires no new
assumption.  More generally, if all physical phenomena could be reduced to
pointlike coincidences of classical point particles and electromagnetic
{\it rays}, then the hypothesis of locality would be exactly valid. 
However, an electromagnetic {\it wave} has intrinsic scales of length and
time characterized by its wavelength $\lambda$ and period $\lambda/c$. 
For instance, the measurement of the frequency of the wave necessitates
observation of a few oscillations before a reasonable determination can
become possible.  If the state of the observer does not change
appreciably over this period of time, then the hypothesis of locality
would be essentially valid.  This criterion may be expressed as
$\lambda/{\cal L}<<1$, where ${\cal L}$ is the relevant acceleration
length of the observer.  That is, the observer has intrinsic scales of
length ${\cal L}$ and time ${\cal L}/c$ that characterize the degree of
variation of its state.  For instance, ${\cal L} = c^2/a$ for an observer
with translational acceleration $a$, while ${\cal L}=c/\Omega$ for an
observer rotating with frequency $\Omega$ \cite{[1],[2]}.  

The consistency of these ideas can be seen in the case of an accelerating
charged particle.  Imagine a particle of mass $m$ and charge $q$ moving
under the influence of an external force ${\bf F}_{ext}$.  The particle
radiates electromagnetic waves that have a characteristic wavelength
$\lambda\sim {\cal L}$, where ${\cal L}$ is the acceleration length of the
particle.  Thus the interaction of the particle with the electromagnetic
field violates the hypothesis of locality since $\lambda/{\cal L}\sim 1$. 
The radiating charged particle is therefore not momentarily equivalent to an
otherwise identical comoving inertial particle.  This agrees with the
fact that in the nonrelativistic approximation the Abraham-Lorentz
equation of motion of the particle is 

\begin{equation}
m\frac{d{\bf v}}{dt} - \frac{2}{3}\frac{q^2}{c^3}\frac{d^2{\bf v}}{dt^2} +
... = {\bf F}_{ext}\;\;,\label{1}
\end{equation}
so that the state of a radiating particle is not determined by
its position and velocity alone.  

Imagine an accelerated measuring device in Minkowski spacetime.  The
internal dynamics of the device is then subject to inertial effects that
consist of the inertial forces of classical mechanics together with their
generalizations to electromagnetic and quantum domains.  If the net
influence of these inertial effects integrates --- over the relevant
length and time scales of a measurement --- to perturbations that do not
appreciably disturb the result of the measurement and can therefore be
neglected, then the hypothesis of locality is valid and the device can be
considered standard (or ideal).  Consider, for instance, the measurement
of time dilation in terms of muon lifetime by observing the decay of
muons in a storage ring.  It follows from the hypothesis of locality that
$\tau_{\mu} = \gamma\tau^0_{\mu}$, where $\gamma$ is the Lorentz factor
and $\tau_{\mu}^0$ is the lifetime of the muon at rest in the background
inertial frame.  On the other hand, the lifetime of such a muon has been
calculated on the basis of quantum theory by assuming that the muon
occupies a high-energy Landau level in a constant magnetic field \cite{[3]}.  One
can show that the result of \cite{[3]} can be expressed as \cite{[4]}

\begin{equation}
\tau_{\mu} \simeq \gamma\tau^0_{\mu} \left[ 1 +
\frac{2}{3}\left(\frac{\lambda}{\cal L}\right)^2\right]\;\;.\label{2}
\end{equation}
Here $\lambda = \hbar/(mc)$ is the Compton wavelength of the
muon, $m$ is the muon mass and ${\cal L} = c^2/a$, where
$a=\gamma^2v^2/r$ is the effective centripetal acceleration of the muons
in the storage ring.  The hypothesis of locality is completely adequate
for such experiments since $\lambda/{\cal L}$ is extremely small.  In
fact, the hypothesis of locality is clearly valid in many Earth-bound
experimental situations since $c^2/g_{\oplus}\simeq 1\:{\rm lyr}$ and
$c/\Omega_{\oplus}\simeq 28\:{\rm AU}$. 

The hypothesis of locality plays a crucial role in Einstein's theory of
gravitation:  Einstein's principle of equivalence together with the
hypothesis of locality implies that an observer in a gravitational field
is locally inertial.  Indeed, the equivalence between an observer in a
gravitational field and an accelerated observer in Minkowski spacetime is
useless operationally unless one specifies what an accelerated observer
measures.  

The hypothesis of locality was formally introduced in \cite{[1]} and its
limitations were pointed out.  To clarify the origin of this conception,
some background information is provided in section 2.  The implications
of this assumption for length determination in rotating systems are
pointed out in section 3.  Section 4 contains a discussion. 

\section{Background}
Maxwell's considerations regarding optical phenomena in moving systems
implicitly contained the hypothesis of locality \cite{[5]}.  The fundamental
form of Maxwell's theory of electromagnetism, derived from Maxwell's
original electrodynamics of media, is essentially due to Lorentz's
development of the theory of electrons.  

Lorentz conceived of an electron as an extremely small charged particle
with a certain smooth volume charge density.  A free electron at rest was
regarded as a spherical material system with certain internal forces that
ensured the constancy of its size and form.  An electron in translational
motion would then be a flattened ellipsoid according to Lorentz, since it
would be deformed from its original spherical shape by the
Lorentz-FitzGerald contraction in the direction of its motion.  The
internal dynamics of electrons therefore became a subject of scientific
inquiry and in 1906 Poincar\'{e} postulated the existence of a particular
type of internal stress that could balance the electrostatic repulsion
even in a moving (and hence flattened) electron.  These issues are
discussed in detail in the fifth chapter (on optical phenomena in moving
bodies) of Lorentz's book \cite{[6]} on the theory of electrons.  

In extending the Lorentz transformations in a pointwise manner to
accelerating electrons, Lorentz encountered a problem regarding the
dynamical equilibrium of the internal state of the electron.  To avoid
this problem, Lorentz introduced a basic assumption that is discussed in
section 183 of his book \cite{[6]}: 

\begin{quote}
``... it has been presupposed that in a curvilinear motion the electron
constantly has its short axis along the tangent to the path, and that,
while the velocity changes, the ratio between the axes of the ellipsoid
is changing at the same time.''  
\end{quote}

To elucidate this assumption, Lorentz explained its approximate validity as
follows (\S 183 of \cite{[6]}): 

\begin{quote}
``... If the form and the orientation of the electron are determined by
forces, we cannot be certain that there exists at every instant a state
of equilibrium.  Even while the translation is constant, there may be
small oscillations of the corpuscle, both in shape and in orientation,
and under variable circumstances, i.e. when the velocity of translation
is changing either in direction or in magnitude, the lagging behind of
which we have just spoken cannot be entirely avoided.  The case is
similar to that of a pendulum bob acted on by a variable force, whose
changes, as is well known, it does not instantaneously follow.  The
pendulum may, however, approximately be said to do so when the variations
of the force are very slow in comparison with its own free vibrations. 
Similarly, the electron may be regarded as being, at every instant, in the
state of equilibrium corresponding to its velocity, provided that the
time in which the velocity changes perceptibly be very much longer than
the period of the oscillations that can be performed under the influence
of the regulating forces.''
\end{quote}
It is therefore clear that the hypothesis of locality and its
limitation were discussed by Lorentz for the case of the motion of
electrons. 

Einstein, in conformity with his general approach of formulating
sym\-metry-like principles that would be independent of the specific nature
of matter, simply adopted the same general assumption for rods and
clocks.  In fact, in discussing the rotating disk problem, Einstein
stated in a footnote on page 60 of \cite{[7]} that:

\begin{quote}
``These considerations assume that the behavior of rods and clocks
depends only upon velocities, and not upon accelerations, or, at least,
that the influence of acceleration does not counteract that of velocity.''
\end{quote}
The modern experimental foundation of Einstein's theory of
gravitation necessitates that this assumption be extended to all (standard)
measuring devices; therefore, the hypothesis of locality supersedes the
clock hypothesis, etc.  

Though the hypothesis of locality originates from Newtonian mechanics,
one should point out that the state of a relativistic point particle
differs from that in Newtonian mechanics: the magnitude of velocity is
always less than $c$.  Moreover, the hypothesis of locality rests on the
possibility of defining instantaneous inertial rest frames along the
worldline of an arbitrary point particle.  In fact, Minkowski raised this
possibility and hence the corresponding hypothesis of locality to the level
of a fundamental axiom \cite{[8]}.  

Another aspect of Lorentz's presupposition must be mentioned here that
involves the extension of the notion of rigid motion to the relativistic
domain:  the electron moves rigidly as it is always undeformed in its
momentary rest frame.  The notion of rigid motion in the special and
general theories of relativity has been discussed by a number of authors
\cite{[9],[10],[11],[12],[13]}.  It is important to note that the concept of an infinitesimal
rigid rod is indispensable in the theory of relativity (cf. section 3).  

In some expositions of relativity theory, such as \cite{[10]} and \cite{[14]}, the
hypothesis of locality is completely implicit.  For instance, in
Robertson's paper on ``Postulate {\it versus} Observation in the Special
Theory of Relativity'' \cite{[14]}, attention is simply confined to ``the
kinematics {\it im kleinen} of physical spacetime'' \cite{[14]}.  However, when
interpreting the observational foundations of special relativity, one must
recognize that actual observers are all accelerated and that the
difference between accelerated and inertial observers must be
investigated; in fact, this problem is ignored in \cite{[14]} by simply asserting
that physics is essentially local.  

\section{Length measurement}

To illustrate the nature of the hypothesis of locality, it is interesting
to consider spatial measurements of rotating observers.  Imagine
observers $A$ and $B$ moving on a circle of radius $r$ about the origin
in the $(x, y)$-plane of a background global inertial frame with
coordinates $(t, x, y, z)$.  Expressed in terms of the azimuthal angle
$\varphi$, the location of $A$ and $B$ at $t=0$ can be chosen such that
$\varphi_A = 0$ and $\varphi_B = \Delta$ with no loss in generality.  The
motion of $A$ and $B$ is then assumed to be such that for $t>0$ they
rotate in {\it exactly the same way} along the circle with angular
frequency ${\hat\Omega}_0(t)>0$.  Thus for $t>0$ observers $A$ and $B$
can be characterized by the azimuthal angles

\begin{equation}
\varphi_A(t) =
\int^t_0\;{\hat\Omega}_0(t^{\prime})dt^{\prime}\;\;,\;\;\varphi_B(t) =
\Delta + \int^t_0\;{\hat\Omega}_0(t^{\prime}) dt^{\prime}\;\;.\label{3}
\end{equation}

According to the static inertial observers in the background global
frame, the angular separation of $A$ and $B$ is constant at any time
$t>0$ and is given by $\varphi_B(t) - \varphi_{A}(t) = \Delta$; moreover,
the spatial separation of the two observers along the circular arc at
time $t>0$ is $\ell(t) = r\Delta$.  

Consider now a class of observers $O$ populating the whole arc from $A$
to $B$ and moving exactly the same way as $A$ and $B$.  At any time
$t>0$, it appears to inertial observers at rest in the background frame
that these rotating observers are all at rest in the $(x^{\prime},
y^{\prime}, z^{\prime})$ system that is obtained from $(x, y, z)$ by a
simple rotation about the $z$-axis with frequency ${\hat\Omega}_0(t)$. 
What is the length of the arc according to these rotating observers?  It
follows from an application of the hypothesis of locality that for
$t>0$ the spatial separation between $A$ and $B$ as measured by the
rotating observers is $\ell^{\prime} = {\hat\gamma}\ell(t)$, where
${\hat\gamma}$ is the Lorentz factor corresponding  to ${\hat v} =
r{\hat\Omega}_0(t)$.  Units are chosen here such that $c=1$ throughout
this section.  Indeed at any time $t>0$ in the inertial frame, each observer
$O$ is momentarily equivalent to a comoving inertial observer and the
corresponding infinitesimal element of the arc $\delta\ell$ has a rest
length $\delta\ell^{\prime}$ in the momentarily comoving inertial frame
such that from the Lorentz transformation between this local inertial
frame and the global background inertial frame one obtains 

\begin{equation}
\sqrt{1-{\hat v}^2}\;\;\delta\ell^{\prime} = \delta\ell\;\label{4}
\end{equation}
in accordance with the Lorentz-FitzGerald contraction. 
Defining 

\begin{equation}
\ell^{\prime} = \Sigma\;\delta\ell^{\prime}\;\;,\label{5}
\end{equation}
where each $\delta\ell^{\prime}$ is the infinitesimal length at
rest in a different local inertial frame, one arrives at
$\ell^{\prime} = {\hat\gamma}\ell$, since ${\hat v}(t)$ is the same for
the class of observers $O$ at time $t$.  The same result is obtained if
length is measured using light travel time over infinitesimal distances
between observers $O$, since in each local inertial frame the two methods
give the same answer.  As is well known, the light signals could also be
used for the synchronization of standard clocks carried by observers
$O$.  

It is important to remark here that equation (\ref{5}) is far from a proper
geometric definition of length and one must question whether it is even
physically reasonable, since each $\delta\ell^{\prime}$ in equation (\ref{5})
refers to a different local Lorentz frame.  In any case, in this approach
the length of the arc as measured by the accelerated observers is 

\begin{equation}
\ell^{\prime} = {\hat\gamma}(t) r\Delta\;\;.\label{6}
\end{equation}  
 
The sum in equation (\ref{5}) involves infinitesimal rest segments each from a
separate local inertial frame.  Perhaps the situation could be improved
by combining these infinite disjoint local inertial rest frames into one
continuous {\it accelerated} frame of reference.  The most natural way to
accomplish this would involve choosing one of the noninertial observers
on the arc and establishing a geodesic coordinate system along its
worldline.  In such a system, the measure of separation along the
worldline (proper time) and away from it (proper length) would also be
determined by the hypothesis of locality.  That is, at any instant of
proper time the rules of Euclidean geometry are applicable as the
accelerated observer is instantaneously inertial.  It turns out that the
length of the arc determined in this way would in general be different
from $\ell^{\prime}$ and would depend on which reference observer $O:
A\rightarrow B$ is chosen for this purpose \cite{[15]}.  To illustrate this
state of affairs and for the sake of concreteness, in the rest of this
section the length of the arc will be determined in a geodesic coordinate
system along the worldline  of observer $A$ and the result will be
compared with equation (\ref{6}).  

In the background inertial frame, the coordinates of observer $A$ are 

\begin{equation} x^{\mu}_A = (t, r\cos\varphi_A, r\sin\varphi_A,
0)\;\;,\label{7} 
\end{equation}
and the proper time along the worldline of $A$ is given by 

\begin{equation}
\tau =\int^{t}_{0}\;\sqrt{1-{\hat v}^2(t^{\prime})}\;\;dt^{\prime}\;\;,
\label{8}\end{equation}
where $\tau=0$ at $t=0$ by assumption.  It is further
assumed that $\tau=\tau(t)$ has an inverse and the inverse function is
denoted by $t=F(\tau)$.  Thus $dt/d\tau = dF/d\tau = \gamma(\tau) =
(1-v^2)^{-1/2}$ is the Lorentz factor along the worldline of $A$, so that
$v(\tau):={\hat v}(t)$ and $\gamma(\tau):={\hat\gamma}(t)$.  Moreover, it
is useful to define $\phi(\tau):=\varphi_A(t)$ and
$d\phi/d\tau=\gamma\Omega_0(\tau)$, where
$\Omega_0(\tau):={\hat\Omega}_0(t)$.  With these definitions, the natural
orthonormal tetrad frame along the worldline of $A$ for $\tau>0$ is given
by

\begin{eqnarray}
\lambda^{\mu}_{{\:\:}(0)} &=&\gamma(1, -v\sin\phi, v\cos\phi,
0)\;\;,\label{9}\\
\lambda^{\mu}_{{\:\:}(1)} &=& (0,\cos\phi, \sin\phi,
0)\;\;,\label{10}\\
\lambda^{\mu}_{{\:\:}(2)} &=& \gamma(v, -\sin\phi, \cos\phi, 0)\;\;,\label{11}\\
\lambda^{\mu}_{{\:\:}(3)} &=& (0\;,\;0\;,\;0\;,\;1)\;\;,\label{12}
\end{eqnarray}
 where $\lambda^{\mu}_{{\:\:}(0)} = dx^{\mu}_A/d\tau$ is the
temporal axis and the spatial triad corresponds to the natural spatial
frame of the rotating observer.  To obtain this tetrad in a simple fashion,
first note that by setting $r=0$ and hence $v=0$  and $\gamma = 1$ in
equations (\ref{9}) - (\ref{12}) one has the natural tetrad of the fixed noninertial
observer at the spatial origin --- as well as the class of noninertial
observers at rest in the background inertial frame --- that refers its
observations to the axes of the $(x^{\prime}, y^{\prime}, z^{\prime})$
coordinate system alluded to before; then, boosting this tetrad with speed
$v$ along the second spatial axis tangent to the circle of radius
$r$ results in equations (\ref{9}) - (\ref{12}).  

It follows from the orthonormality of the tetrad system (\ref{9}) - (\ref{12}) that
the acceleration tensor ${\cal A}_{\alpha\beta}$ defined by 

\begin{equation}
\frac{d\lambda^{\mu}_{{\:\:}(\alpha)}}{d\tau} = {\cal
A}_{\alpha}^{\;\;\beta}\lambda^{\mu}_{{\:\:}(\beta)}\label{13}
\end{equation}
is antisymmetric.  The translational acceleration of observer
$A$, which is the ``electric'' part of the acceleration tensor $(a_i =
{\cal A}_{0i})$, is given by 

\begin{equation} {\bf a} = (-\gamma^2
v\Omega_0\;\;,\;\;\gamma^2\frac{dv}{d\tau}\;\;,\;\;0)\label{14}
\end{equation}
with respect to the tetrad frame and similarly the rotational
frequency of $A$, which is the ``magnetic'' part of the acceleration
tensor $(\Omega_i = \frac{1}{2}\epsilon_{ijk}{\cal A}^{jk})$, is given by 

\begin{equation}
\mbox{\boldmath$\Omega$} = (0, 0, \gamma^2\Omega_0)\;\;.\label{15}
\end{equation}
Moreover, in close analogy with electrodynamics, one can define
the invariants of the acceleration tensor as

\begin{equation} I = -a^2+\Omega^2 = \gamma^2\Omega^2_0 -
\gamma^4\left(\frac{dv}{d\tau}\right)^2\label{16}
\end{equation}
and $I^{*} = -{\bf a}\cdot\mbox{\boldmath$\Omega$} = 0$.  The
analogue of a null electromagnetic field is in this case a null
acceleration tensor; that is, an acceleration tensor is null if both $I$ and
$I^{*}$ vanish.  A rotating observer with a null acceleration tensor is
discussed in the appendix.  

The translational acceleration ${\bf a}$ consists of the well-known
centripetal acceleration $\gamma^2 v^2/r$ and the tangential acceleration
$\gamma^2 dv/d\tau$.  The latter formula is consistent with the
corresponding result in the case of linear acceleration along a fixed
direction.  To interpret equation (\ref{15}) as the frequency of rotation of
the spatial frame with respect to a local nonrotating frame, it is
necessary to construct a nonrotating, i.e. Fermi-Walker transported,
orthonormal tetrad frame ${\tilde\lambda}^{\mu}_{{\:\:}(\alpha)}$ along the
worldline of observer $A$.  Let ${\tilde\lambda}^{\mu}_{{\:\:}(0)}=
\lambda^{\mu}_{{\:\:}(0)}\;\;,\;\;{\tilde\lambda}^{\mu}_{{\:\:}(3)} =
\lambda^{\mu}_{{\:\:}(3)}$ and

\begin{eqnarray}
{\tilde\lambda}^{\mu}_{{\:\:}(1)} & = & \cos\Phi\;\lambda^{\mu}_{{\:\:}(1)}
-\sin\Phi\;\lambda^{\mu}_{{\:\:}(2)}\;\;,\label{17}\\
{\tilde\lambda}^{\mu}_{{\:\:}(2)} & = & \sin\Phi\;\lambda^{\mu}_{{\:\:}(1)}
+\cos\Phi\;\lambda^{\mu}_{{\:\:}(2)}\;\;,\label{18}
\end{eqnarray}
where the angle $\Phi$ is defined by 

\begin{equation}
\Phi = \int^{\tau}_0\;\Omega(\tau^{\prime}) d\tau^{\prime}\;\;,\label{19}
\end{equation}
so that $d\Phi/d\tau = \gamma^2\Omega_0$.  It remains to show
that ${\tilde\lambda}^{\mu}_{{\:\:}(i)}\;,\;i=1, 2, 3$, correspond to local
ideal gyroscope directions.  This can be demonstrated explicitly using
equations (\ref{17}) - (\ref{19}) and one finds that 

\begin{equation}
\frac{d\tilde{\lambda}^{\mu}_{{\:\:}(i)}}{d\tau} =
\tilde{a}_i\;{\tilde{\lambda}}^{\mu}_{{\:\:}(0)}\;\;,\label{20}
\end{equation}
where ${\bf\tilde{a}}$ is the translational acceleration with
respect to the nonrotating frame, as expected.  It is straightforward to
study the average motion of the spatial frame
${\tilde{\lambda}}^{\mu}_{{\:\:}(i)}$ with respect to the background
inertial axes and illustrate Thomas precession with frequency
$(1-{\hat\gamma}){\hat\Omega}_0$ per unit time $t$.  That is, the frame of
the accelerated observer rotates with frequency ${\hat\Omega}_0(t)$ about
the background inertial axes, while the Fermi-Walker transported frame
rotates with frequency
$-{\hat\gamma}{\hat\Omega}_0$ per unit time $t$ with respect to the frame
of the accelerated observer according to equations (\ref{17}) - (\ref{19}); therefore,
the unit gyroscope directions precess with respect to the background
inertial frame with frequency $(1-{\hat\gamma}){\hat\Omega}_0$ as
measured by the static background inertial observers.  

Along the worldline of observer $A$, the geodesic coordinates can be
introduced as follows:  At a proper time $\tau$, consider the straight
spacelike geodesics that span the hyperplane orthogonal to the
worldline.  An event $x^{\mu} = (t, x, y, z)$ on this hyperplane is assigned
geodesic coordinates $X^{\mu} = (T, {\bf X})$ such that 

\begin{equation}
x^{\mu} = x^{\mu}_A(\tau) + X^i\lambda^{\mu}_{{\:\:}(i)}(\tau)\;\;,\;\;\tau
= T\;\;.\label{21}
\end{equation}
Let ${\bf X} = (X, Y, Z)$ and recall that along the worldline of
$A, t = F(\tau)$ and $\varphi_A(t) = \phi(\tau)$; then, the transformation
to the new coordinates is given by 

\begin{eqnarray}
t &=& F(T) + \gamma(T) v(T) Y\;\;,\label{22}\\
x &=& (X+r)\cos\phi(T) - \gamma(T) Y\sin\phi(T)\;\;,\label{23}\\
y &=& (X+r)\sin\phi(T) + \gamma(T) Y\cos\phi(T)\;\;,\label{24}\\
z &=& Z\;\;.\label{25}
\end{eqnarray}
For $r=0$, the geodesic coordinate system reduces to
$(t^{\prime}, x^{\prime}, y^{\prime}, z^{\prime})$, where $t^{\prime} = t$;
that is, the standard rotating coordinate system is simply the geodesic
coordinate system constructed along the worldline of the noninertial
observer at rest at the origin of spatial coordinates.  

The form of the metric tensor in the geodesic coordinate system has been
discussed in \cite{[1],[15],[16]}.  It turns out that in the case under
consideration here the geodesic coordinates are admissible within a
cylindrical region \cite{[16]}.  The boundary of this region is a real elliptic
cylinder for $I>0$, a parabolic cylinder for $I=0$ or a hyperbolic
cylinder for $I<0$, where the acceleration invariant $I$ is given by
equation (\ref{16}).  

The class of observers $O: A\rightarrow B$ lies on an arc of the circle
$x^2+y^2=r^2$ in the background coordinate system; therefore, it follows
from equations (\ref{23}) and (\ref{24}) that in the geodesic coordinate system the
corresponding figure is an ellipse

\begin{equation}
\frac{(X+r)^2}{r^2} + \frac{Y^2}{(r\gamma^{-1})^2} = 1\label{26}
\end{equation}
with semimajor axis $r$, semiminor axis $r\sqrt{1-v^2}$ and
eccentricity $v$.  The latter quantities are in general dependent upon time
$T$, hence at a given time $t$ each observer lies on a different ellipse. 
It is natural to think of the ellipse (\ref{26}) as a circle of radius $r$ that
has suffered Lorentz-FitzGerald contraction along the direction of motion
\cite{[1],[15]}.  

The measurement of the length from $A$ to $B$ in the new system involves
the integration of $dL, dL^2 = dX^2 + dY^2$, along the curve from $A:
(T_A, 0, 0, 0)$ to $B: (T_B, X_B, Y_B, 0)$ corresponding to $A: (t,
r\cos\varphi_A$, $r\sin\varphi_A, 0)$ and $B: (t, r\cos\varphi_B,
r\sin\varphi_B, 0)$ in the background inertial frame.  To clarify the
situation, it is useful to introduce --- in analogy with the elliptic
motion in the Kepler problem --- the eccentric anomaly $\theta$ by 

\begin{equation}
X+r = r\cos\theta\;\;,\;\;Y = r\sqrt{1-v^2}\sin\theta\;\;.\label{27}
\end{equation}
Then, for a typical rotating observer $O: (t, r\cos\varphi,
r\sin\varphi, 0)$ on the arc from $A\rightarrow B$ with 

\begin{equation}
\varphi = \delta + \int^t_0\;{\hat\Omega}_0(t^{\prime})dt^{\prime}\label{28}
\end{equation}
one has in geodesic coordinates $O : (T, X, Y, 0),$ where $X$ and
$Y$ are given by equations (\ref{27}), and equations (\ref{22}) - (\ref{24}) imply that 

\begin{equation}
t = F(T) + r v(T)\sin\theta\;\;,\label{29}
\end{equation}

\begin{equation}
\varphi = \theta + \phi(T)\;\;.\label{30}
\end{equation}
As $O$ ranges from $A$ to $B$, $\delta:0\rightarrow\Delta$ in
equation (\ref{28}) and hence $\theta: 0 \rightarrow\Theta$.  For a fixed $t$, $t =
F(T_A),$ equation (\ref{29}) can be solved to give $T$ as a function of $\theta$;
then, a detailed calculation involving equations (\ref{27}) - (\ref{30}) shows that 

\begin{equation}
L = r\int^{\Theta}_0\;\sqrt{1-v^2W\cos^2\theta}\;d\theta\;\;.\label{31}
\end{equation}
Here $W$ is defined by 

\begin{equation}
W = \gamma^2\;\frac{1 - r^2\dot{v}^2\sin^2\theta}{(\gamma +
r\dot{v}\sin\theta)^2}\;\;,\label{32}
\end{equation}
$\dot{v} = dv/dT$ and $\Theta$ can be found in terms of $\Delta$
by solving equations (\ref{29}) and (\ref{30}) at $B$: 

\begin{equation}
t = F(T_B) + rv(T_B)\sin\Theta\;\;,\label{33}
\end{equation}

\begin{equation}
\Delta + \int^{t}_{0}\;{\hat{\Omega}}_0 (t^{\prime}) dt^{\prime} = \Theta +
\phi(T_B)\;\;.\label{34}
\end{equation}

In practice, the explicit calculation of $L$ can be rather complicated;
therefore, for the sake of simplicity only the case of constant $v$ (i.e.
uniform rotation) will be considered further here \cite{[1],[15]}.  Then, $W=1$
and equation (\ref{31}) simply refers to the arc of a constant ellipse for
which a Kepler-like equation

\begin{equation}
\theta - v^2\sin\theta = \delta\label{35}
\end{equation}
follows from equations (\ref{29}) and (\ref{30}).  Furthermore, the proper
acceleration length of the uniformly rotating observer $A$ is given by
${\cal L} = I^{-1/2} = (\gamma\Omega_0)^{-1}$.  The case of uniform
rotation, where $L$ and $\ell^{\prime}$ are independent of time and
$L\neq\ell^{\prime}$ in general,  has been treated in detail in \cite{[1],[15]} and
it is clear that irrespective of the magnitude of
$\Delta, L/\ell^{\prime}\rightarrow 1$ as $r/{\cal L} = v\gamma\rightarrow
0$; on the other hand for $\Delta\rightarrow 0, L/\ell^{\prime}\rightarrow
1$ irrespective of $v<1$.  That is, consistency can be achieved only if the
length under consideration is negligibly small compared to the
acceleration length of the observer.  

\section{Discussion}
It is important to recognize that the hypothesis of locality is an
essential element of the theories of special and general relativity.  In
particular, it is indispensable for the measurement of spatial and
temporal intervals by accelerated observers.  Therefore, relativistic
measurement theory must take this basic assumption and its limitations
into account.  This has been done for the measurement of time in \cite{[17]}. 
In connection with the measurement of distance, it has been shown that
there is a lack of uniqueness; however, this problem can be resolved if
the distance under consideration is much smaller than the relevant
acceleration length of the observer \cite{[1],[15]}.  This means that from a
basic standpoint the significance of noninertial reference frames is
rather limited \cite{[16]}.  In practice, however, the difference between $L$
and $\ell^{\prime}$ (discussed in section 3) is usually rather small; for
instance, in the case of the equatorial circumference of the Earth this
difference amounts to about $10^{-2}\:{\rm cm}$ \cite{[15]}.  

The application of these concepts to standard accelerated measuring
devices that are by definition consistent with the hypothesis of locality
results in a certain maximal acceleration \cite{[18],[19]} that is imposed by the
quantum theory.   For a classical device of mass $M$, the dimensions of
the device must be much larger than $\hbar/(Mc)$ according to the quantum
theory of measurement \cite{[20],[21]}.  On the other hand, the dimensions of the
device must be much smaller than its acceleration length ${\cal L}$ .  It
follows that ${\cal L}>>\hbar/(Mc)$ for any standard classical measuring
device \cite{[2],[4]}.  Thus for ${\cal L} = c^2/a$, the translational
acceleration $a$ must be much smaller than $Mc^3/\hbar$, while for ${\cal
L}=c/\Omega$, the rotational frequency $\Omega$ must be much smaller than
$Mc^2/\hbar$.  Further discussion of the notion of maximal acceleration
is contained in \cite{[22]}.  

The hypothesis of locality is compatible with wave phenomena only when
the latter are considered in the ray limit $(\lambda/{\cal L}\rightarrow
0)$.  To go beyond the basic limitations inherent in the hypothesis of
locality regarding the treatment of wave phenomena, a nonlocal theory of
accelerated observers has been developed \cite{[23],[24],[25]}.  In this theory, the
amplitude of a radiation field as measured by an accelerated observer
depends on its history, namely, its past worldline in Minkowski
spacetime.  This acceleration-induced nonlocality constitutes the first
step in the program of developing a nonlocal theory of gravitation.   

\appendix{Null acceleration}
The relativistic theory of an observer in arbitrary circular motion is
treated in section 3.  In this case, the proper acceleration length of the
observer is defined to be $|I|^{-1/2}$, where $I$ is given by equation
(\ref{16}).  It is interesting to study the circular motion of an observer with a
constant prescribed magnitude of $I$.  In fact, equation (\ref{16}) can be
written as 

\begin{eqnarray}
\left(\frac{d{\hat v}}{dt}\right)^2 = \frac{1}{r^2}\;{\hat v}^2(1-{\hat
v^2})^2 - I(1-{\hat v}^2)^3 \nonumber\;,
\end{eqnarray}
which for constant $I$ can be simply integrated.  For the null
acceleration case $I=0$, the solution is

\begin{eqnarray}
{\hat v}^{-2} = 1 + \eta\;e^{\mp 2\frac{t}{r}}\nonumber
\end{eqnarray}
for $\eta>0$.  The upper sign refers to motion
that asymptotically $(t\rightarrow\infty)$ approaches the speed of light,
while the lower sign corresponds to an asymptotic state of rest.

\end{document}